# Physical basis of critical analysis.

Yu.S. Tsyganov

*FLNR, JINR. 141980 Dubna, Russia*

*tyura@sungns.jinr.ru*

**Abstract**

*A new approach to provide a fruitful analysis of signal amplitude is proposed. It is this method that has made it possible to eliminate some observed artificial events from the list of candidates for decays of superheavy nuclei. Examples of estimates of measured amplitudes of evaporation residues (EVR) are presented. Some attention is paid to a new "hard" statistical criterion for detecting rare events. This criterion can be applied in the case of significant role of random factors.*

1. Introduction

With the discovery of the "island of stability" of superheavy nuclei by the Dubna Gas Filled Recoil Separator group [1-3] a few different questions can be considered as an actual matter. One of them is related to establishing of the edges of the "island". Recently, some attempts to start those experiments were performed at Dubna and Darmstadt. Another interesting question of a different matter is: how perfect, fast and reliable should be the detection system for such experiments. Application of digital system on the base of DSSSD detector together (in parallel) with the modern analog system with a purpose of search for pointer to Recoil-alpha correlation in a real time mode can be considered as a reasonable scenario [4, 5]. Note, that sometimes in analysis of sets of experimental data it is useful to apply not only the knowledge about predicted properties of the nuclides under investigation and values of reaction cross sections, but some knowledge of "non nuclear" nature related to applications of silicon radiation detectors. In some cases this analysis can give indication of negative nature, that is, it indicates to a low quality of the given event interpretation.

2. Application of silicon radiation detectors in the experiments aimed at the synthesis of SHE.

In the last years in the experiments aimed at the synthesis of SHE's at different facilities in FLNR (JINR, Dubna), GSI (Darmstadt, Germany), Riken (Saitama, Japan), LBNL (Berkeley, USA), PSI ( Villigen, Switzerland) position sensitive silicon radiation detectors were the main part of the detection systems. Namely with these devices it has become possible to establish a genetic link between individual alpha decays of superheavy nuclei. Of course, a strong statistical analysis is required to avoid an interpretation based on a random factor explanation [6,7 ]. To this end namely knowledge about silicon radiation detectors allows experimentalists to reject some events from the list of potential candidates.  Below, a few examples of amplitude signal analysis are given.

## 2.1 Detection of highly ionized EVR's signals of implanted heavy nuclei from heavy-ion – induced nuclear reactions

It is a well known fact that detection of EVR signal is very important from the viewpoint of general experimental philosophy. Mostly, due to an evident fact that namely from that signal starts a whole multi chain event and, therefore, it allows to estimate half life value for a product under investigation. Another well known fact, that under detection with silicon radiation detector amplitude of registered signal for strongly ionizing particle can be presented in the form of equation:

$E_{REG}$=E – PHD. Here $E_{REG}$ – is a registered energy signal, E- incoming energy value and PHD – **P**ulse **H**eight **D**efect value (of course, after a definite calibration procedure with alpha particle source is performed by the experimentalists).

PHD value is composed from three components, namely: losses in the entrance window, recombination component and nuclear stopping one [8].

For the DGFRS detection system in [9] an approximate relation $E_{REG}$= $E_{REG}$(E) has been derived in the form of equation: $E_{REG} \approx -1.7 + 0.74 \cdot E_{IN}$ [ MeV's, 10 < E < 40].

Additionally, not only empirical relations are useful, but any Monte Carlo calculations too.

In [10] such approach is performed. With these approach in [11] was shown that one from three detected events of Z=112 element has an artificial nature (Fig.1a).

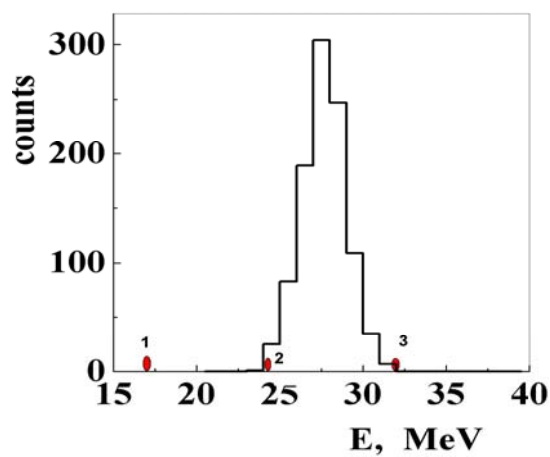

**Fig.1 a**  Event "1" is considered as having no relation to the effect. It was eliminated from t5he event list by the authors

On the other hand, Fig.1 b,c demonstrates good agreement for EVR events detected and for Z=118 events detected in FLNR(JINR) and correction (systematic error) function (b). In Fig.2 results for Z=112 EVR's measured in GSI for complete fusion nuclear reaction $^{238}$U+ $^{48}$Ca→$^{283}$Cn + 3n are shown.

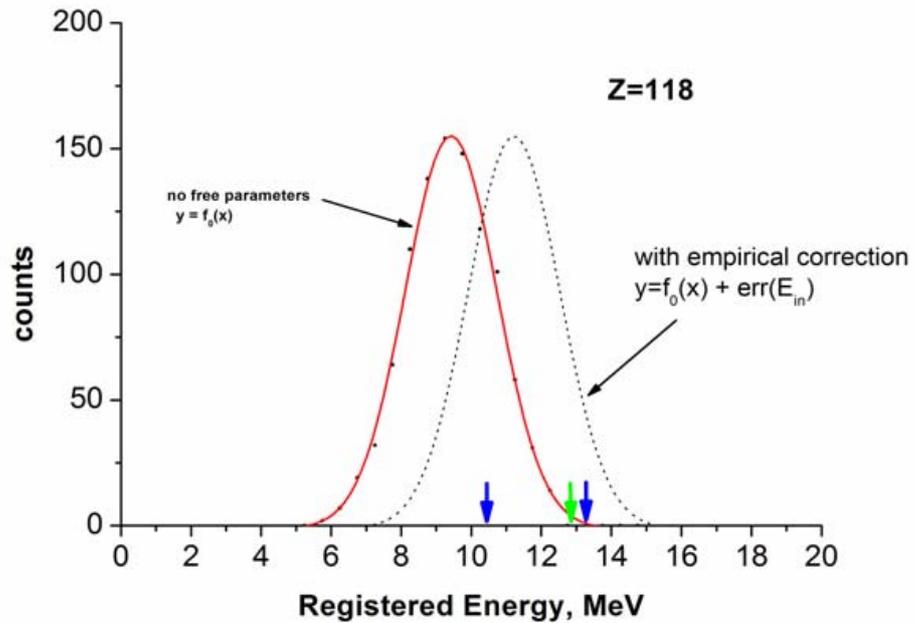

**Fig.1 b** PC simulations for Z=118 (dotted line-correction to error function)

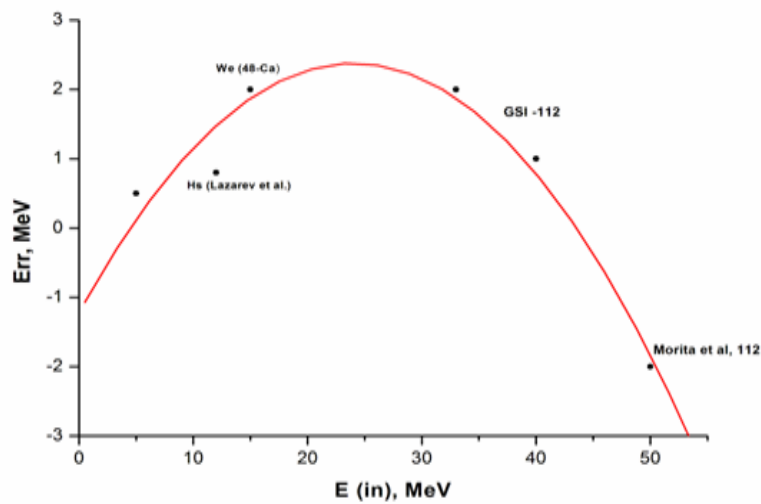

**Fig.1 c** Dependence of calculation values error against EVR's incoming energy value

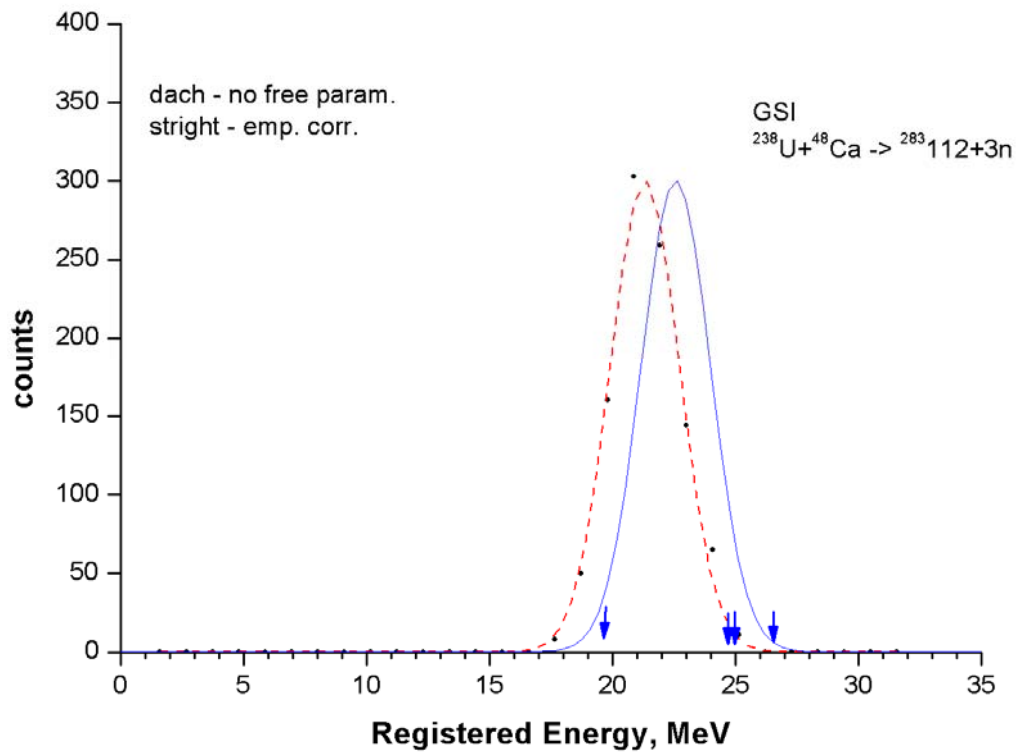

**Fig.2** GSI Z=112 EVR's four events. Dash line- no free parameters of simulation. Straight line – program calibration error function is taken into account.

**2.2 Detection of spontaneous fission fragments of implanted nuclei**

Very often, multi chain event contains spontaneous fission signal as a finishing one. Because of implantation depth of EVR in silicon is of several microns some specific exists in the process of those signals detection.  Although the calibration of SF scale is usually quite a delicate process, definite systematic is applied for experimental result analysis if to use dimensionless parameters. In [12] parameter $k = \frac{E(esc)}{E(foc)+E(esc)}$ is proposed to detect any deviations from systematic, where indexes (esc) and (foc) are corresponded to backward and focal plane detectors, respectively.

In Fig.3 those parameters are shown for different cases are indicated.  It can be easily seen, that energy in the experiment (areas 1, 2 - data from Ref. [13] ) was overestimated and, probable, true values of total kinetic energy are slighter smaller.

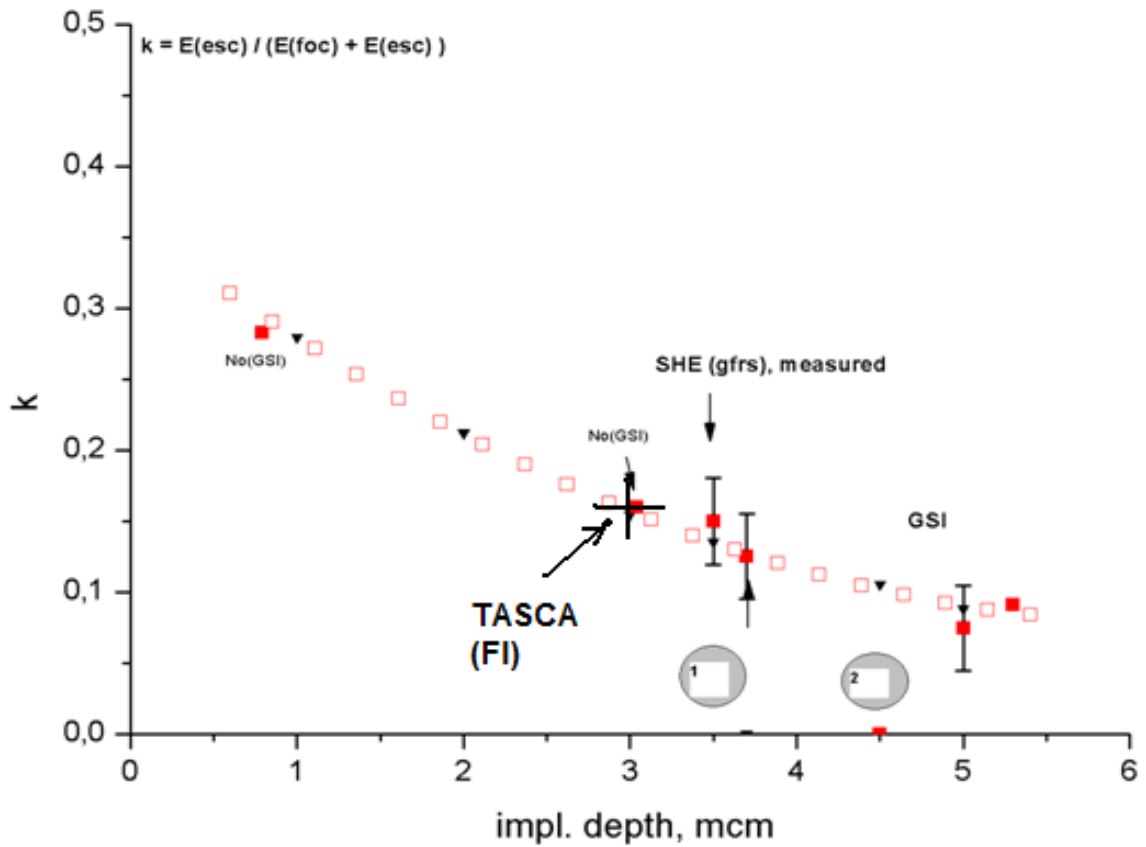

**Fig.3** The dependence of k-parameter against implantation depth. Areas 1,2 indicates to Z=115 TASCA experiment [13] Rectangles – calculated values. Area 1 – implantation depth calculated from EVR energies, 2- from kinematic calculation starting from EVR energy in the middle of the target.

Additionally, the mean measured value of EVR's energy in the experiment [14] is:

$<E_{EVR}>$ = <13.4, 13.9, 16.3, 16.1, 16.4, 15.4, 14.1, 12.5, 15.9, 14.5, 16.5, 15.1, 15.3, 14.9, 17.2 > ≈ 15.2 MeV.

But, according to the above mentioned formula (with $E_{in}^{EVR}$≈33.4 MeV)

$<E_{CALC}>$≈-1.7 +33.4·0.74 = 23 MeV.

Probable, that discrepancy is explained by thicker entrance window (more than ~1µm Si) DSSSD detector in [13] with respect to PIPS detectors were used for obtaining the empirical formula.

## 2.3 A few words about statistics

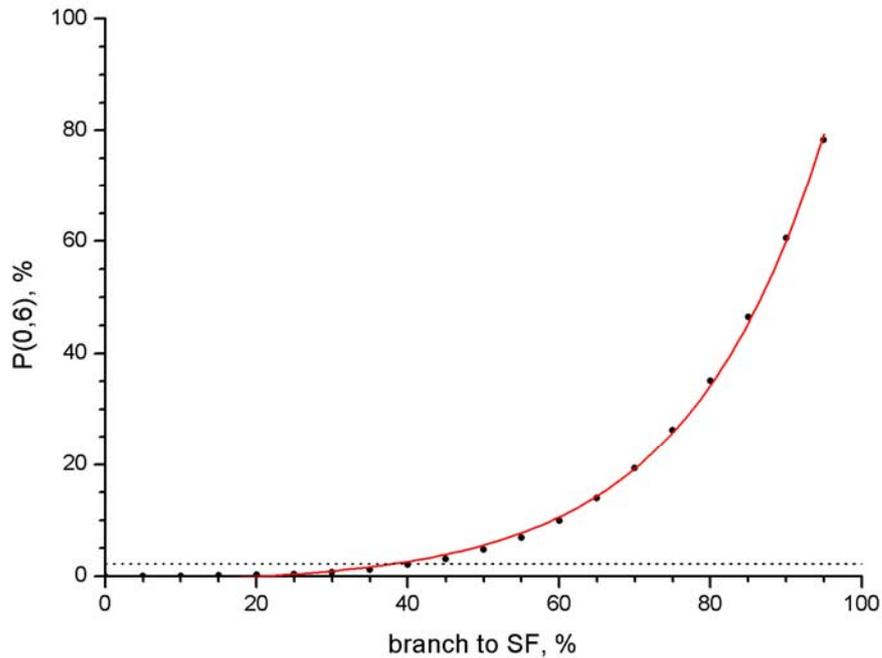

**Fig.4** Probability P(0,6) for six missed alpha particle signals against branch to spontaneous fission value

Sometimes it is useful to calculate probability of missing of several numbers of alpha-particle chains when one detects event like EVR-α-SF. In Fig.4 parameter of probability to miss alpha particle signal for six measured events against the branch to spontaneous fission for mother nuclide is shown. Dotted line indicates level which was presented in Ref. [14].

Another point of view, related to statistics, is the following: calculated (estimated) parameter of expectation for random events $N_R$ is sometimes (if relation of $N_R<<1$ is not too valid, i.e., has no excess of several orders of magnitude)it is useful to apply more strongest criteria with normalization that parameter onto the probability of the given chain configuration [15 ].

Namely:

$$\tilde{N}_R = N_R/P_{CONF},$$

where $P_{CONF}$ is a configuration probability.

In that case a criterion for a true event is $\tilde{N}_R << 1$ and for a true background event - $\tilde{N}_R >> 1$. All other area of $\tilde{N}_R$ is declared as "*non clear event*" or "*non true background*". Namely this estimate was made for one Z=114 event candidate reported in [16].

### 3. Specifics in application of DSSSD detector

When applying DSSSD detector one should definitely bear in mind a significant role of an edge effect (charge dividing between neighbor strips), especially for back side (with respect to beam direction) strips. It strongly required to take into account that percentage of those effects is depend on the source geometry (Fig.5a,b).

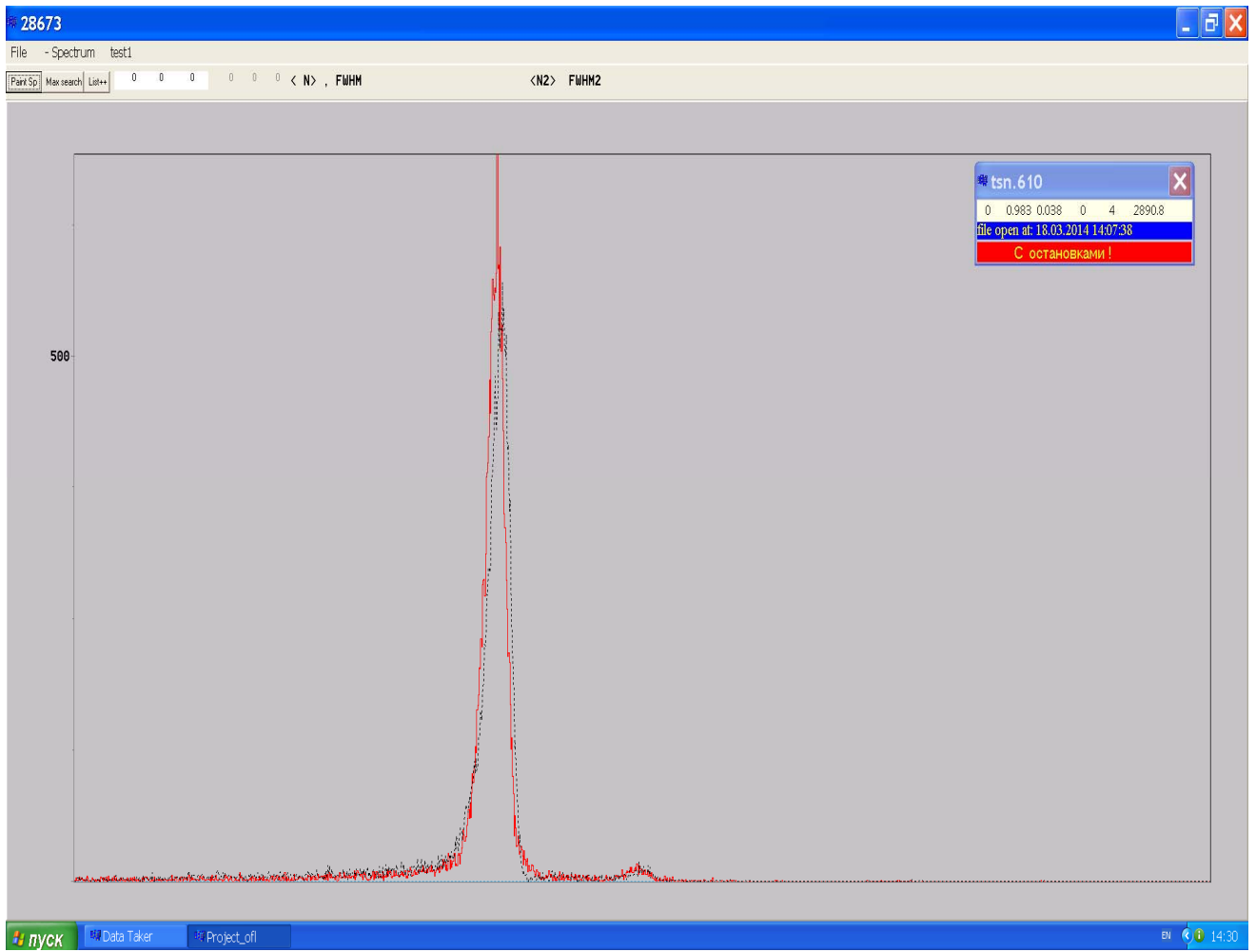

Fig.5 a Summary alpha spectra measured with front 48 strips; (line) and back 128 strips (dotted line) of DSSSD detector. Value of 0.038 (right upper corner) indicates part of signals measured with two neighbor strips. (On-line measurements)

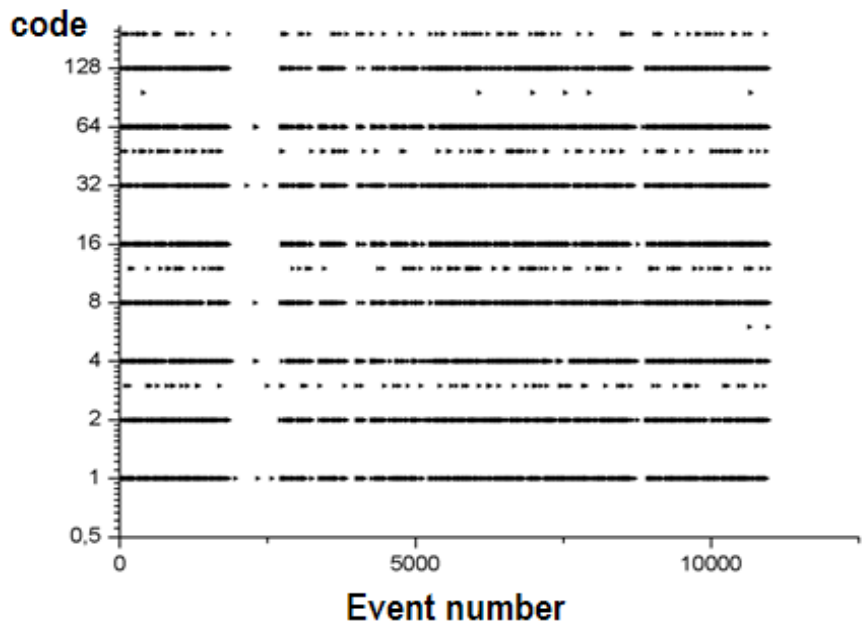

Fig.5b  Edge effects for back strips of DSSSD detector demonstration. Dotted lines- both (neighbor) strips signal sharing. Codes $2^n$ – single strip operations.

That is, when for normally incoming ~5 MeV alpha particle signals a value for both (neighbor) strip signals may be about 3-4%, whereas for geometry, close to $2\pi$ one can achieves up to 20%. Therefore, when one detects a few alpha decays of implanted nuclei, a value close to 20% should be for signals from neighbor strips (for the DGFRS DSSSD detector manufactured by Micron Semiconductor).

**4 Summary**

In conclusion, typical schematics can be proposed for critical analysis of events ascribed to superheavy nuclei. The principal steps are:

- To simulate measured EVR energy spectra and to compare with the empirical systematic for nuclei with the closest (Z, A). If necessary to do re-calibration of simulation code;
- To compare measured events of SHE with both systematics (semi-empirical and/or simulated ones);
- To compare dimensionless k-parameter for measured SF events with calculated one;

    If parameter of $N_R$ is not too small, to use more "hard" criterion to eliminate artificial events;

- If one declares losses of some alpha-particle signals it is necessary to estimate probability of those given multi event configuration;
- Estimation of the relative value of number of signals which can be explained by strip-to-strip edge effect is desirable too.

Author is indebted to Drs. A.N. Polyakov, A.A. Voinov and A.B. Yakushev for their help in preparing of this paper and fruitful discussion.

The paper is supported in part by the RFBR Grant №13-02-12052.

## 5. Supplement 1

When preparing this manuscript, new paper confirming the DGFRS result on the properties of $^{294}117$ isotope has been published [17]. Two decay chains were reported. These chains start with ER signals of 6.9 and 9.0 MeV, respectively. Taking into account an extra dead layer of TASCA DSSSD detector of about 0.5 µm (Si) [18], one can state a good agreement with the spectrum simulated in [12] (see, e.g. Fig.2 too).

## 6. Supplement 2

Of course, it is possible to consider re-calibration procedure taking into account mass difference between calibration isotopes and measured ones (e.g. [19]).

Let us consider an "ideal" calibration case, namely:

K = $e_0/N_0$, where k-calibration constant, $e_0$-energy yield related with electron-hole pairs generation in silicon, $N_0$- channel number.

If one takes into account that $e_0 = e_0^\alpha \cdot (1 + \eta \cdot \frac{m}{M_0})$, where $M_0 \approx 217\text{-}4$ ($^{217}$Th alpha decay from $^{nat}$Yt +$^{48}$Ca reaction) and m=4, η-part of energy yield for electron-hole generation.

Therefore, when detecting SHE alpha decay and under assumption the same η value mostly explaining by the stopping component of PHD (pulse height defect for heavy low energy recoil) one can write: $e_{SHE} = e_{SHE}^\alpha \cdot (1 + \eta \cdot \frac{m}{M_{SHE}})$.

Additionally, assuming the new channel $N_{SHE}$ is in the close vicinity of $N_0$ that is N1/N0 ~ 1.

Finally, it is possible to obtain relation for alpha particle energy in the form of:

$\mu = \frac{e_{SHE}^\alpha}{e_0^\alpha} \approx \frac{1+\eta \cdot \frac{m}{M_0}}{1+\eta \cdot \frac{m}{M_{SHE}}}$.  Author considers the above mentioned formulae as a reasonable scenario to estimate a systematic error of the calibration process.

For M=294-4 and Z=117 (e≈11 MeV) η≈0.3 ([19]), hence, correction factor µ is approximately equal to :

$$\mu = \frac{1+0.3\cdot 4/(217-4)}{1+0.3\cdot 4/(294-4)} \approx 1.0015.$$ In the vicinity of 11 MeV region it creates roughly the measured energy shift value as 11000·(µ-1)≈17 KeV.

Note, that for more precise estimation the recombination component of PHD should be taken into account, as well as the difference for η-parameter for masses $M_0$ and $M_{SHE}$.